\documentclass[preprint]{elsarticle}

\usepackage{amsmath}
\usepackage{graphicx}
\usepackage{natbib}

\newcommand{\dvec}[1]{\ensuremath{\boldsymbol{#1}}}
\newcommand{\vk}{\dvec{\mathrm{k}}}
\newcommand{\vecr}{\dvec{\mathrm{r}}}

\newcommand{\eV}{\ensuremath{\mathrm{eV}}}

\begin{document}

\title{Compressibility of graphene}
\author{D. S. L. Abergel}
\address{Condensed Matter Theory Center, Department of Physics,
University of Maryland, College Park, MD 20742, USA}

\begin{abstract}
	We present a review of the electronic compressibility
	of monolayer and bilayer graphene.
	We focus on describing theoretical calculations of the effects of
	electron--electron interactions and various types of disorder, and
	also give a summary of current experiments and describe which
	aspects of theory they support.
	We also include a full analysis of all commonly-used contributions
	to the tight-binding Hamiltonian of bilayer graphene and their
	effects on the compressibility.
\end{abstract}

\maketitle

\section{Introduction}

The compressibility of the electron liquid, $\kappa$, is a fundamental
physical quantity which can yield information about the interactions
within the fluid and the effect of extrinsic influences such as
disorder. In the clean, non-interacting limit, the compressibility is
straightforwardly expressed in theory in terms of quantities which can
be easily derived from the band structure. 
Also, the compressibility is experimentally accessible since it is
related to the quantum capacitance of the electron liquid, $C_Q$, and
can be measured directly by single electron transistor mounted on a
scanning probe microscope.
Both the compressibility and the quantum capacitance may be computed
from $d\mu/dn$, where $\mu$ is the chemical potential and $n$ is the
carrier density relative to charge neutrality, since 
\begin{equation}
	\frac{1}{\kappa} = n^2 \frac{d\mu}{dn} \quad\text{and}\quad
	C_Q = \mathcal{A} e^2 \frac{dn}{d\mu}
\end{equation}
where $\mathcal{A}$ is the sample area. Therefore the study of this
quantity $M\equiv d\mu/dn$ is of high importance in condensed matter
physics and is the principal subject which we will address in this
review.

Monolayer graphene and its bilayer have become the center of an intense
amount of research because of their novel properties, the potential
for advances in the fundamental understanding of massless and massive
chiral electrons, and for the many ways in which these materials might be
applied in the creation of devices. 
In this paper, we review the current state of knowledge of the
compressibility and related quantities in these materials. We focus on
theoretical descriptions of the effect of disorder,
electron--electron interactions, and the fundamental properties of the
electron liquid, but we also discuss the current experimental
data and describe which parts of the theory these measurements support.
The range of features is more rich in bilayer graphene so we shall
inevitably devote more attention to this topic, but this is not to
diminish from the importance of understanding the situation in monolayer
graphene.

To accomplish this, in the remainder of this introduction, we describe
the pertinent single particle properties of monolayer graphene and its
AB-stacked bilayer, including the derivation of $M$ in the clean,
non-interacting limit.
Then in Section \ref{sec:elelint} we review the role of
electron--electron interactions and in Section \ref{sec:disorder} we
describe the modifications due to various types of disorder. In Section
\ref{sec:exp} we compare these various theories to current experimental
data and finally, in Section \ref{sec:conclusion}, we summarize the
results and discuss their implications.

\begin{figure}
	\centering
	\includegraphics[]{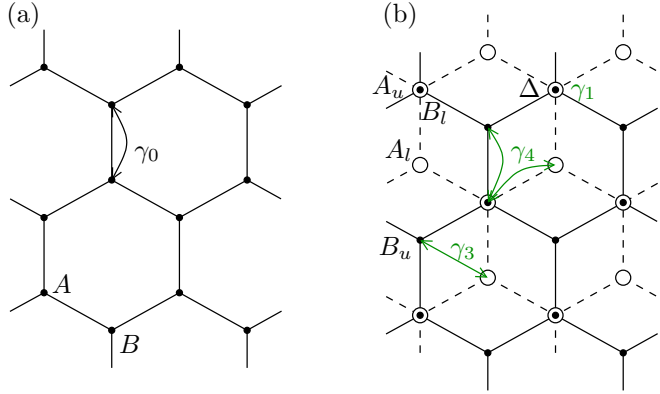}
	\caption{(Color online.) The lattice structure of (a) monolayer and
	(b) bilayer graphene annotated with the various hopping elements
	which are included in the tight-binding Hamiltonian. The dashed
	lines and larger unfilled circles denote the lower layer, the solid
	lines and
	smaller filled circles show the upper layer. Intra-layer and onsite
	elements are shown in black, inter-layer hops are drawn in green
	(gray).
	\label{fig:hopping}}
\end{figure}

We use the tight-binding formalism to describe the properties of
electrons in graphene. 
Many thorough introductions to this theory already exist
\cite{abergel-advphys59, dassarma-rmp83, castroneto-rmp81}, but we shall
describe in detail the features which we shall utilize in this review to
ensure that all notations and conventions are properly defined.
In the case of the monolayer, there are two inequivalent lattice sites
in the unit cell, which we label $A$ and $B$ sites as shown in
Fig.~\ref{fig:hopping}(a). The lattice constant is
denoted $a$ and the nearest-neighbor (NN) distance is $a/\sqrt{3}$.
The electron hopping process between these two sublattices is
characterized by the energy $\gamma_0$ which yields the monolayer Fermi
velocity $v_F = \sqrt{3} a \gamma_0 / (2\hbar) \approx
10^6\mathrm{ms}^{-1}$. 
We exclude next-nearest neighbor hops because they do not produce any
substantial change to the band structure, and we do not include any
sublattice asymmetry since there is no well-controlled way of
implementing this in an experimental context.

The Hamiltonian of monolayer graphene in one valley may be written in
leading order in momentum as
\begin{equation}
	H^m = \xi \hbar v_F \dvec{\sigma}\cdot \vk =
	\begin{pmatrix} 0 & \xi \hbar v_F \pi^\dagger \\
	\xi \hbar v_F \pi & 0 \end{pmatrix}
\end{equation}
where the sublattice basis is $\{A,B\}$ in the K valley (with $\xi=1$) 
and $\{B,A\}$ in the K$^\prime$ valley (with $\xi=-1$). 
The operator $\pi = k_x + i k_y$ is the linear expansion of the
transition matrix elements near the K points.
The energy spectrum associated with this Hamiltonian is
\begin{equation}
	E_{\nu k}^m = \nu \hbar v_F k
\end{equation}
where $\nu = \pm1$ denotes the conduction and valence bands and $k =
|\vk|$.

The crystal structure of bilayer graphene consists of two monolayer
lattices stacked such that the $A$ sublattice of the upper layer lies
directly above the $B$ sublattice of the lower layer, separated by a
distance $a_z$. 
This is called Bernal, or AB stacking and is illustrated in
Fig.~\ref{fig:hopping}(b). A dimer bond is formed between these two
lattices and is characterized by the energy $\gamma_1$ in the
tight-binding Hamiltonian. 
In the bilayer, the sublattice asymmetries and next-nearest inter-layer
hops do have a significant effect on the band structure and we describe
the most relevant ones now. Firstly the layer potential asymmetry, which
we denote by the energy $u$, breaks the inversion symmetry in the
out-of-plane direction and lifts the degeneracy of the conduction and
valence bands at the K points. Since this potential can be applied by
gating, this generates a dynamically-tunable band gap
\cite{mccann-prl96, mccann-prb74}. 
The hops between $A_l$ and $B_u$ lattices, parameterized by the energy
$\gamma_3$ break the isotropy of the band structure reducing the
symmetry from full rotational to C$_3$, and induce an overlap between the
conduction and valence bands. The hops between $A_l$ and $A_u$
sublattices and between $B_l$ and $B_u$ sublattices parameterized by the
energy $\gamma_4$ combine with the
intra-layer potential asymmetry $\Delta$ to induce an asymmetry between
the conduction and valence bands. 

\begin{figure}
	\centering
	\includegraphics[]{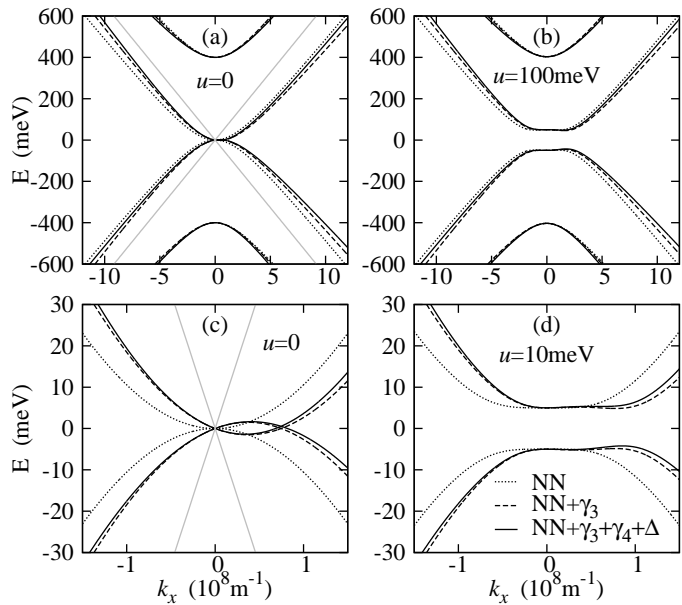}
	\caption{The bandstructure of bilayer graphene. (a) $u=0$
	showing all four bands. (b) $u=100\mathrm{meV}$ showing all four
	bands. (c) Low-energy region for $u=0$. (d) Low-energy region for
	$u=10\mathrm{meV}$. The legend in (d) applies to all four panels,
	and `NN' stands for nearest-neighbor. The gray line in (a) and (c)
	is the dispersion of monolayer graphene.
	\label{fig:bs}}
\end{figure}

Using these notations, the Hamiltonian is
\begin{equation}
	H^b = \begin{pmatrix} \tfrac{\xi u}{2} & 
		\xi \hbar v_3 \pi^\dagger &
		-\xi \hbar v_4 \pi^\dagger &
		\xi \hbar v_F \pi^\dagger \\
	\xi \hbar v_3 \pi &
		-\tfrac{\xi u}{2} &
		\xi \hbar v_F \pi &
		- \xi \hbar v_4 \pi \\
	-\xi \hbar v_4 \pi &
		\xi \hbar v_F \pi^\dagger &
		-\tfrac{\xi u}{2} + \Delta &
		\gamma_1 \\
	\xi \hbar v_F \pi &
		-\xi \hbar v_4 \pi^\dagger &
		\gamma_1 &
		\tfrac{\xi u}{2} + \Delta
	\end{pmatrix}
	\label{eq:Hbil}
\end{equation}
where the sublattice basis is $\{B_u, A_l, B_l, A_u\}$ with $\xi=1$ in
the K valley and $\{A_l, B_u, A_u, B_l\}$ with $\xi=-1$ in the
K$^\prime$ valley. The energies $\gamma_3$ and $\gamma_4$ give the
velocities $v_{3,4} = \sqrt{3}\gamma_{3,4} a / (2\hbar)$.
Figure \ref{fig:bs} shows the band structure of bilayer graphene in
three approximations. Solid lines denote the full solution of
Eq.~\eqref{eq:Hbil} with $\gamma_0=3.09\eV$, $\gamma_1=0.4\eV$,
$\gamma_3=0.38\eV$, $\gamma_4=0.14\eV$, and $\Delta=0.022\eV$ where
these values have been taken so that $v_F=10^6\mathrm{ms}^{-1}$ and
$\gamma_1$ is in line with standard values taken in the literature. The
higher order parameters were taken from an experimental study which
extracted their values by fitting optical reflectivity spectra to the
band structure of the full Hamiltonian \cite{kuzmenko-prb80}.
Dashed lines indicate the spectrum when the asymmetry-inducing elements
$\gamma_4$ and $\Delta$ are set to zero; and the dotted lines show the
spectrum when the trigonal warping elements $\gamma_3$ are also set to
zero. This final approximation is the most frequently used in the
literature. Figure \ref{fig:bs}(a) shows the hyperbolic shape of
the bands in the gapless case, and the asymmetries induced by the
higher order tight binding parameters. Specifically, the trigonal
warping terms push the bands towards positive $k_x$, and while it is not
shown in the figure, this asymmetry causes the Fermi surface to become
somewhat triangular, reducing its full isotropy to a three-fold
rotational symmetry. The combined effect of $\gamma_4$ and $\Delta$ is
to make the conduction band slightly steeper and the valence band
somewhat shallower. We also include the linear dispersion of monolayer
graphene for comparison, shown with the gray line.
Figure \ref{fig:bs}(b) shows the same spectrum but
with a gap generated by $u=100\mathrm{meV}$. The `sombrero' shape near
the K point is now noticeable, as is the distortion caused by the higher
order tight binding terms which make the gap slightly narrower at the
positive $k_x$ minimum. Figure \ref{fig:bs}(c) shows the low-energy
region for the gapless case, and the overlap of the conduction and
valence bands caused by $\gamma_3$ is now apparent. Figure
\ref{fig:bs}(d) shows the same but for a small gap given by
$u=10\mathrm{meV}$. Most of the complexity of the low energy band
structure is encapsulated in this plot, including the distortion of the
band edge and all the asymmetries described above.

There is a common simplification of the bilayer Hamiltonian which is
often used when only the very low density regime is under investigation
\cite{mccann-prl96}. By expanding the Green's function of the
Hamiltonian in Eq.~(\ref{eq:Hbil}) with respect to momentum, one can
write down an effective two band Hamiltonian with quadratic dispersion.
This effectively excludes the $A_u$ and $B_l$ sublattices from
consideration, since electrons residing primarily on these sublattices
form the split bands with energy $\sim \gamma_1$ near the K points and
which are therefore not involved in low energy processes.

In order to derive exact expressions for $M$, we must first find an
analytical expression for the density in terms of the chemical
potential. Elementary considerations show us that the total number of
electrons in the conduction band (or holes in the valence band) is
related to the number of filled states as follows: 
\begin{align*}
	N = \sum_{\text{filled. st.}} 1 
	&= \frac{g_s g_v \mathcal{A}}{(2\pi)^2} \int f(\vk) d^2 \vk \\
	\Rightarrow \qquad
	n &= \frac{g_s g_v}{4\pi^2} \int_0^{2\pi} d\theta
	\int_{k_{F_-}(\theta)}^{k_{F_+}(\theta)} k \, dk
\end{align*}
where $n=N/\mathcal{A}$, and the state occupancy at zero temperature is
given by the function $f(\vk)$ which takes the value 1 if the state with
wave vector $\vk$ is occupied and 0 otherwise. 
The possible anisotropy and non-trivial topology of the
band structure implies that the Fermi wave vector may be
multiply-valued, and is a function of the angle $\theta$. The
factors $g_s$ and $g_v$ are for spin and valley degeneracies,
respectively, and both take the numerical value of 2.
Evaluating the radial part gives
\begin{equation}
	n = \frac{g_s g_v}{8\pi^2} \int_0^{2\pi} d\theta \left(
	k_{F_+}^2(\theta) - k_{F_-}^2(\theta) \right)
	\label{eq:nintegral}
\end{equation}
This integral equation (which depends on the chemical potential via the
Fermi wave vectors) can now be evaluated to give an expression for $\mu$
in terms of $n$. This can then be differentiated with respect to $n$
to give $M$.

This process can be carried out analytically in a number of cases. For
monolayer graphene, the Fermi surface is always circular such that
$k_{F_-}(\theta) = 0$ and $k_{F_+}(\theta) = k_F$. Then,
Eq.~(\ref{eq:nintegral}) becomes
\begin{equation}
	n = \frac{g_s g_v}{8\pi^2} \int_0^{2\pi} d\theta \left(
	k_F^2 - 0 \right) = \frac{k_F^2}{\pi} \quad\Rightarrow\quad
	k_F^2 = \pi n.
\end{equation}
The band structure also gives us $\mu = \hbar v_F k_F = \hbar v_F
\sqrt{\pi n}$ so that 
\begin{equation} 
	M^m = \frac{\hbar v_F \sqrt{\pi}}{2\sqrt{n}}.
\end{equation}
Similarly, the quadratic band structure yields a constant value for $M$:
\begin{equation}
	M^q = \frac{\hbar^2 v_F^2 \pi}{\gamma_1}.
\end{equation}
The four-band model for bilayer graphene with $\gamma_3 = \gamma_4 =
\Delta = 0$ also gives an analytical result. 
However, in this case, the Fermi energy may be ring-shaped for low
density, and for high density both the low-energy and split bands are
populated. Therefore $M$ shows two step-like features as the Fermi
energy moves between these regions:
\begin{equation}
	M^h = \frac{\gamma_1 u}{\sqrt{g_u}} \delta(n) +
	\frac{\hbar^2 v_F^2 \pi}{2} \begin{cases}
	\frac{\lambda}{\sqrt{g_u}} \frac{1}{\sqrt{\lambda^2 +
	u^2\gamma_1^2}} & \lambda < u^2 \\
	\frac{ 1 - g_u / \left(2 \sqrt{\lambda g_u + \gamma_1^4/4} \right)
		}{\sqrt{\lambda + \frac{u^2}{4} + \frac{\gamma_1^2}{2}
	- \sqrt{\lambda g_u + \frac{\gamma_1^4}{4}} }} 
	& u^2 \leq \lambda < 2\gamma_1^2 + u^2 \\
	\frac{1}{\sqrt{2\lambda + u^2}} & \lambda \geq 2\gamma_1^2 + u^2
	\end{cases}
\end{equation}
where we use the notation $g_u = \gamma_1^2 + u^2$ for brevity. The
first term of this expression which contains the $\delta$-function is
present because in the gapped case, a discontinuity exists in the
chemical potential at the band edge: for an infinitesimal change in
density, the chemical potential jumps from the valence band to the
conduction band, and differentiating this step gives the $\delta$
function as shown.
Including the higher-order tight binding elements means that analytical
solutions are no longer possible, and numerical evaluation must be
conducted instead. 

\begin{figure*}
	\centering
	\includegraphics[]{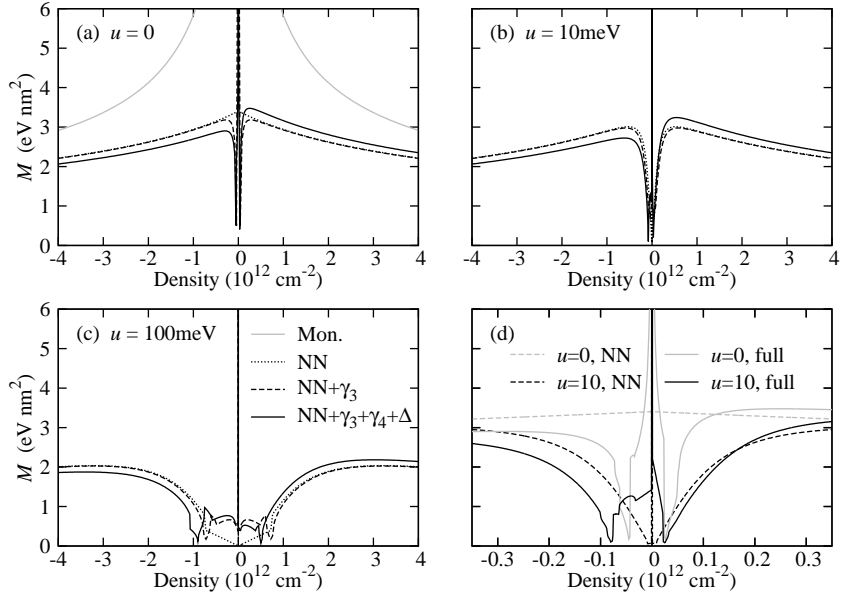}
	\caption{$M=d\mu/dn$ for clean bilayer graphene.
	(a) The gapless case, also with $M^m$ for monolayer graphene.
	(b) $u=10\mathrm{meV}$, and (c) $u=100\mathrm{meV}$. The legend in
	(c) shows the approximation for the bilayer Hamiltonian used in each
	case and applies to panels (a)--(c).
	(d) Low-density region for the full Hamiltonian and three different
	gap values (corresponding to the solid lines from the other three
	panels).
	\label{fig:cleandmudn}}
\end{figure*}

Figure \ref{fig:cleandmudn} shows $M$ in the various cases we have
described. We point out the important features. Firstly, in the gapless
case, shown in Fig.~\ref{fig:cleandmudn}(a), the monolayer (gray line)
$M$ is substantially larger than the bilayer and diverges as $1/\sqrt{n}$ as
$n\to 0$. In the NN approximation for the bilayer, $M$
goes smoothly to the finite value $\hbar^2 v_F^2 \pi/\gamma_1$ at $n=0$.
The decrease in $M$ at finite density reflects the non-parabolicity of
the band structure. When the second-order hopping elements are included,
a sharp peak is present at $n=0$ with dips at small density on either
side. This is due to the change in topology of the Fermi surface with
finite $\gamma_3$ and the region where $M$ decreases relative to the NN
case corresponds to the density range where Fermi surface has split into
four pockets. At high density, the asymmetry induced by the combination
of $\gamma_4$ and $\Delta$ manifests as a reduction in $M$ for the
valence band and enhancement for the conduction band. This also reveals
that the presence or absence of $\gamma_3$ from the Hamiltonian has
little effect at higher density. 
When a small gap is opened, shown in Fig.~\ref{fig:cleandmudn}(b) for
$u=10\mathrm{meV}$, the divergence in the density of states at the band
edge manifests as $M\to0$ when $n\to0$.
However, we illustrate the $\delta$-function spike as an
infinitesimally wide divergence exactly at $n=0$. The effect of the
trigonal warping is now masked by the band gap and the asymmetry between
the conduction and valence bands is unmodified. When the gap is wider,
shown for $u=100\mathrm{meV}$ in Fig.~\ref{fig:cleandmudn}(c), there is
an extended range of density (approximately
$|n|<10^{12}\mathrm{cm}^{-2}$) where the Fermi energy is in the sombrero
region. For the NN approximation, $M$ is linear, but for the
higher order approximations of the Hamiltonian, the non-isotropic band
structure creates a rather complex shape for $M$.
In Fig.~\ref{fig:cleandmudn}(d) we show the low-density region for the
gapless and gapped ($u=10\mathrm{meV}$) cases, for the NN and full
approximations for the band structure. This shows the dip--peak
structure in the gapless full approximation more clearly, in addition to
the complexity of $M$ in the gapped full approximation.

Finally, we note that this single particle analysis has been used to
compare the quantum capacitance of bulk monolayer graphene to that of
graphene nanoribbons \cite{fang-apl91}. 

\section{Electron--electron interactions \label{sec:elelint}}

We now consider the effect of direct Coulomb interactions between
electrons and correlations on the compressibility.
The first observation \cite{abergel-prb80} is that in the absence of
direct electron--electron interactions, the kinetic energy
part of the correlation (an expression of the fact that electrons are
fermionic) between two electrons behaves qualitatively
differently for monolayer and bilayer graphene. Specifically, the
two-particle correlation cancels exactly for monolayer (leaving only the
contribution from the trivial single particle kinetic energy part)
whereas it remains finite for the bilayer. This is due to the differences in
the chirality of the two electron liquids. This cancellation in the
correlation implies that the compressibility is determined by only the
single particle kinetic energy, as seen in experiments
\cite{martin-natphys4}. In contrast, the bilayer does not show this
cancellation, implying that correlations and the contribution from the
interactions are likely to play a more significant role in that
material.

To take into account the effects of interactions between electrons,
various levels of approximation have been employed. At the Hartree--Fock
level in monolayer graphene \cite{hwang-prl99}, the renormalized $M$
retains the $n^{-1/2}$ dependence of the non-interacting system, but
the value of $M$ is increased relative to the non-interacting electrons.
The overall enhancement in the extrinsic case is approximately 20\% when
the parameters relevant for graphene mounted on an SiO$_2$ substrate are
used. The authors also showed that the renormalization is increased when
the dielectric environment of the graphene is weaker, suggesting that
the enhancement of $M$ will be stronger in suspended graphene.
At the RPA level with finite doping, interactions between electrons
favor states with large chirality \cite{barlas-prl98} which leads to a
suppression of both charge and spin susceptibilities. In this
approximation, the overall sign of the compressibility remains positive
but the interactions reduce it by up to a factor of 2 corresponding to
an enhancement of $M$, depending on the
strength of the coupling constant (which is defined by the environment
of the graphene) and the value of an arbitrarily defined momentum
cutoff. For physically realistic parameters, a suppression of $\kappa
\propto 1/M$ of $15\%$ is predicted. 

In contrast, the introduction of electron--electron interactions at the
Hartree--Fock level to ungapped bilayer graphene \cite{kusminskiy-prl100}
gives a profound change to the predicted inverse compressibility
(proportional to $M$)
because at low carrier density it becomes negative and divergent. 
Just as in the monolayer, the contributions from the intra-band and
inter-band terms are opposite in sign, but in the case of the bilayer at
low density, the inter-band terms (which are negative) are stronger
which leads to a negative divergence as $n\to0$.
For the monolayer, the (negative) inter-band terms are never strong
enough to overcome the (positive) intra-band contribution.
Also, while the two-band model provides a reasonable approximation for
$M$ at low densities, the authors point out that the full four-band
model is required at intermediate and higher densities.
If the RPA approximation to the correlation part of the exchange is
included \cite{borghi-prb82} then the negative divergence is removed
because 
the contribution to the total energy per particle given by the
correlation effects cancels the decreasing contribution from the
exchange implying that $M$ is always positive. The compressibility
$\kappa$ is reduced with respect to the non-interacting case, but this
enhancement of $M$ reduces with increasing carrier density.

Finally, we mention in passing that the compressibility may be a useful
tool for investigating the nature of the ground state of charge-neutral
(intrinsic) bilayer graphene. There have been many theoretical works
(Refs. \cite{vafek-prb81} and \cite{zhang-prb81} are two of the
earliest) which suggest that this system is unstable against transitions
to various broken-symmetry phases. The nematic phase and anomalous Hall
insulator are two of the strongest candidates, and may lead to
distinctive behavior of $M$ and $\kappa$ at low density. In fact, some
experiments \cite{martin-prl105} have already been interpreted in this
manner.

\section{Disorder \label{sec:disorder}}

\subsection{Introduction}
The effect of disorder is known to be crucial for determining many
observable properties of graphene materials, and in many cases may mask
the effects of electron--electron interactions discussed above.
For example, the Dirac point physics can be completely masked by charge
inhomogeneity \cite{martin-natphys4} and in the more dirty samples, the
presence of the externally tunable band gap in bilayer graphene can be
obscured in transport measurements \cite{oostinga-natmat7}. The exact
cause of the charge inhomogeneity (which manifests as `puddles' at low
average carrier density) is still somewhat controversial, so below we
briefly outline the proposed mechanisms and proceed to assume the
existence of the charge inhomogeneity as a phenomenological reality and
discuss its impact on the compressibility. 

Before discussing the effects of puddles, we mention that for monolayer
graphene, the effect of a finite scattering rate between the electrons
and charged impurities, the effect of ripples, and a
contribution from the correlation between electrons have been examined
\cite{asgari-prb77}, and good agreement between the RPA theory and
experimental data was found at higher density. However, at low density
the theory does not modify the single-particle behavior (a divergence as
$n\to 0$) in contrast to the experimental data. 

\subsection{The effect of puddles \label{sec:puddles}}

The `puddles' of carriers were first observed in monolayer
graphene via scanning SET microscopy \cite{martin-natphys4} and
subsequently in STM measurements for monolayer \cite{deshpande-prb79}
and bilayer \cite{rutter-natphys7, deshpande-apl95} graphene.
The conventional wisdom is that these puddles are caused by external
charged impurities (either located in the substrate, at the interface
between the substrate and the graphene, or on top of the graphene) which
have an electric field associated with them. This field produces a
spatially fluctuating potential which the electrons experience as a
shift in the band structure relative to the Fermi energy which in turn
modifies the number of carriers in that locality. This idea has been
used theoretically to predict the size and distribution of these puddles
\cite{rossi-prl101,rossi-prl107,adam-prb84}, and these studies agree
qualitatively with the experimental measurements for both monolayer and
bilayer graphenes.
Recently, another theory has been advanced \cite{gibertini-arXiv1111} which
suggests that the scalar and vector potentials induced by corrugations
in the graphene sheet may also cause puddles to form. The authors use a
continuum elasticity theory for the graphene sheet and self-consistent
Kohn--Sham--Dirac theory to calculate the induced carrier density. They
find that there is no obvious correlation between the corrugations and
the puddle landscape and state that this is due to the non-local
relation between the corrugation-induced scalar potential and the tensor
field due to height fluctuations. Other work \cite{guinea-prb77} has
also hinted at a link between charge inhomogeneity and deformations of
the graphene sheet since the rippling of monolayer graphene on a
substrate gives rise to a gauge field which may cause low-energy Landau
levels to form, accompanied by an increase in $M$ at low density.

A phenomenological theory which accounts for puddles has been developed
\cite{abergel-prb83, abergel-prb84} for incorporating the effects of
disorder in calculations of $M$.  For a measurement technique which
simultaneously samples an area large
enough to encompass several puddles (such as a bulk capacitance probe
which samples the whole area between the gates, or an SET tip which
samples an area with radius $\sim 100\mathrm{nm}$) and which couples
capacitively to the electronic liquid in the graphene, it is sensible to
assume that the measured response is the average over the sampled area. 
The idea is that each area with similar density will act as an
independent capacitor and that $M$ will be set by that density. Hence
the total of all these parallel regions is their average and this
suggests that the areal average can be replaced by an average over the
density distribution $P$.
In order to model this distribution we write the total local density
$n(\vecr) = n_0 + \tilde{n}(\vecr)$ where $n_0$ is the average density which
in an experiment is set by the external gates, and $\tilde{n}(\vecr)$ is
the spatially fluctuating part induced by the disorder potential. We
also assume that the density distribution is described by one 
parameter---a measure of the average fluctuation which we label $\delta
n$. The average value of $M$, which we denote $\bar{M}$ is
\begin{equation}
	\bar{M}(n_0, \delta n, u ) = 
	\int_{-\infty}^\infty M( n, u ) P( n, n_0, \delta n) dn 
\end{equation}
Note that this procedure contains an uncontrolled approximation that
the Bloch states used in the derivation of $M$ persist in the disordered
context and may be used as a basis over which to perform the average.
Since the density inhomogeneity breaks translational invariance, this is
not strictly true.
Also, for monolayer graphene, this procedure is not completely defined
because the $n^{-1/2}$ divergence in $M$ at low density means that the
integral is infinite. However, an artificial cutoff could be introduced
to rectify this.
The $\delta$-function divergence in the gapped bilayer is
integrable and hence provides no technical difficulty.

\begin{figure*}
	\centering
	\includegraphics[]{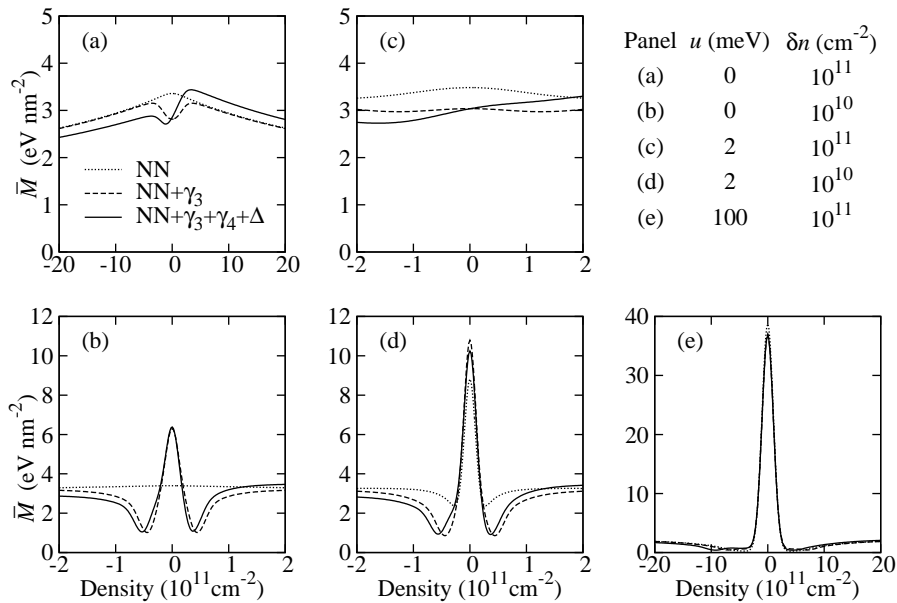}
	\caption{Effect of averaging procedure $\bar{M}$. All lines
	correspond to the legend in (a) (where `NN' stand for
	nearest-neighbor), and the parameters used in each
	panel are shown in the accompanying table.
	\label{fig:Mbar}}
\end{figure*}

In the case of bilayer graphene, this procedure has been used to fit
experimental data with a high degree of accuracy \cite{abergel-prb83,
abergel-prb84} despite the phenomenological nature of the theory. 
To illustrate this, Fig. \ref{fig:Mbar} shows $\bar{M}$ for different
disorder strengths, different values of the gap, and for several
combinations of the tight binding elements $\gamma_3$, $\gamma_4$, and
$\Delta$. 
It is known both experimentally \cite{burson-unpub} and theoretically
\cite{rossi-prl107} that $P$ is Gaussian for bilayer graphene and so we
write
\begin{equation*}
	P( n, n_0, \delta n) = \frac{1}{\sqrt{2\pi} \delta n}
	\exp\left[ -\frac{(n-n_0)^2}{2\delta n^2}
	\right]
\end{equation*}
As representative values of the density fluctuations, we take
$\delta n=10^{11}\mathrm{cm}^{-2}$ to stand for graphene on an SiO$_2$
substrate, and $\delta n=10^{10}\mathrm{cm}^{-2}$ for both suspended
graphene and graphene on an $h$BN substrate.  
According to the literature, these constitute rather conservative
estimates of the density inhomogeneity.
In Fig. \ref{fig:Mbar}(a) we show $\bar{M}$ for strongly
disordered bilayer graphene in a wide density range. At higher density,
the averaging procedure has very little effect and $\bar{M}$ is almost
unchanged from $M$. But at low density the sharp dips in $M$ (see Fig.
\ref{fig:cleandmudn}) become smoothed out into a small dip in $\bar{M}$
near $n=0$. 
When the disorder is much weaker, as shown in Fig.~\ref{fig:Mbar}(b),
the sequence of dips and central peak at $n=0$ is restored (notice the
different scale on the
horizontal axis in this panel). When a small gap is opened, the extra
complexity in $M$ in the low density regime is obscured by the wide
sampling region for strongly disordered samples, as shown in
Fig.~\ref{fig:Mbar}(c).
But for smaller disorder, Fig.~\ref{fig:Mbar}(d), the sequence of dips
on either side of the $n=0$ peak are discernable. In the case of
the wide gap with large disorder, Fig.~\ref{fig:Mbar}(e), the effect of the
second-order tight-binding elements is obscured and the most important
factor determining the shape is the broadened $\delta(n)$ term.

In conclusion, the effect of the puddles on $M$ at low carrier density
in bilayer graphene is an intricate quantitative question in which
two main effects -- the size of the layer asymmetry $u$, and the size of the
density fluctuations -- both play an important role. Additionally, the
shape of $\bar{M}$ is qualitatively different in the gapless case, and
quantitatively different in the presence of a gap, depending on the
approximation taken for the tight-binding Hamiltonian.

\subsection{Spatial fluctuation in band gap}

\begin{figure*}
	\centering
	\includegraphics[]{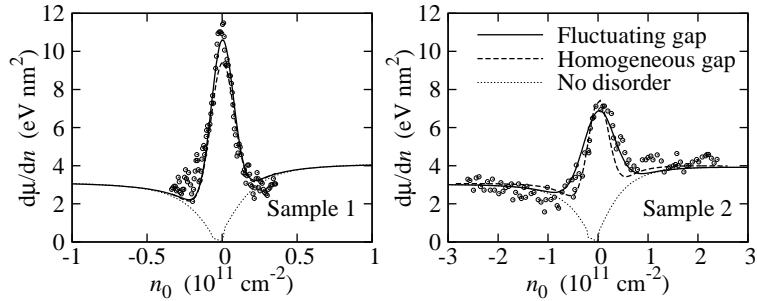}
	\caption{Comparison between gap formation mechanisms in comparison
	to experiments.
	\label{fig:fitcomp}}
\end{figure*}

It has been observed in experiments for graphene
on single-gated SiO$_2$ \cite{rutter-natphys7} that the size of the
inter-layer potential asymmetry (the parameter $u$ in the tight-binding
formalism, which sets the size of the band gap) may fluctuate in space.
Specifically, these authors fitted $dI/dV$ maps of bilayer graphene
in a strong magnetic field to the single particle Landau level spectrum
and were able to extract both the magnitude and sign of $u$. They found
variation on the order of tens of meV, and that the orientation of the
asymmetry swapped between electron puddles and hole puddles. The spatial
fluctuation of the gap was implemented in this phenomenological theory
\cite{abergel-prb84} to model the low-density region of suspended
bilayers by assuming that the local gap could be written $u(\vecr) = u_0
+ \tilde{u}(\vecr)$ in an analogous way to the density fluctuations with
$\delta u$ parameterizing the fluctuations in $\tilde{u}$. Then, a
second averaging procedure can be performed over the distribution of
$\tilde u$ yielding
\begin{equation}
	\bar{\bar{M}} = \int_{-\infty}^\infty \bar{M}(n_0,\delta n,u)
		P(u, u_0, \delta u) du
\end{equation}
It was found that for these suspended samples, it was not possible to
distinguish clearly between a gap formed by spontaneous
electron--electron interaction effects (which is encapsulated in a finite
$u_0$) and a disorder-induced gap modeled by finite $\delta u$. 
This is summarized in Fig. \ref{fig:fitcomp} where the best fits from
the theories for fluctuating gap ($\bar{\bar{M}}$) and homogeneous gap
($\bar{M}$) are compared to experimental data for two samples.
Therefore, if possible, measurements of $M$ in samples with even less
disorder are required to unambiguously determine the magnitude of the
intrinsic gap.

\section{Experiments \label{sec:exp}}

Experimental determinations of $M$ or the compressibility of graphene
have been made via scanning SET microscopy \cite{martin-natphys4,
martin-prl105}, and bulk capacitance measurements \cite{henriksen-prb82,
young-prb84}. 
In the case of monolayer graphene \cite{martin-natphys4}, 
the measured $M$ is close to the single particle result with a Fermi
velocity of $1.1\times10^6 \mathrm{ms}^{-1}$. This corresponds to the
theoretical result where the $1/\sqrt{n}$ divergence persists in the
presence of disorder with a renormalization of the Fermi velocity. The
effects of disorder at low density are noticeable as a cutting off of
the divergence as $n\to 0$.

There have been more experiments done on bilayer graphene. In
Refs.~\cite{henriksen-prb82, young-prb84} $M$ is extracted from bulk
capacitance measurements of dual-gated bilayer sheets in close contact
to the gates.
This geometry induces a large degree of disorder from charged impurities
and the dominant effects shown in these experiments are the broadening
of features due to the puddle averaging described in Section
\ref{sec:puddles}. The asymmetry between the electron and hole bands
caused by $\gamma_4$ and $\Delta$ are also clearly seen in
\cite{henriksen-prb82}.
In contrast, scanning SET microscopy of suspended
bilayer graphene \cite{martin-prl105} accesses much cleaner samples. It
is claimed that the increase in $M$ measured near $n=0$ in these
experiments comes from the formation of a broken symmetry ground state
such as a nematic phase \cite{vafek-prb81} or an anomalous Hall 
insulator \cite{zhang-prb81}. However, other features in the theory
described in Section \ref{sec:elelint} (such as the negative divergence of
$\kappa^{-1}$ in the ungapped bilayer, or quantitative estimates of the
size of the renormalization of $M$ due to electron--electron
interactions) have not been observed.

\section{Conclusions and discussion \label{sec:conclusion}}

The compressibility of an electron liquid and its related quantities
$M=d\mu/dn$ and the quantum capacitance are important objects of study in
condensed matter physics since they give information about the intrinsic
nature of the electron liquid and its interactions with external fields.
In this review, we have described the band structure and detailed
features of $d\mu/dn$ in monolayer and bilayer graphene, including the
effects of electron--electron interactions and disorder. We have also
reviewed the current status of experimental work.  The single particle
picture for $d\mu/dn$ is rather clear for monolayer graphene and for
bilayer graphene at high carrier density. But for the bilayer in the low
density limit, there are qualitative differences between the various
commonly used approximations to the Hamiltonian. In particular, when the
next-nearest neighbor hoppings are included in the gapless case,
significant additional structure appears at experimentally relevant
density.  Comparison with current experiments shows that disorder
(specifically, the existence of inhomogeneity in the charge density
landscape) is the dominant source of deviations from the single particle
picture in graphene samples in contact with a substrate, while in
suspended samples it is possible that the interaction effects are being
observed.

We gratefully acknowledge support from US-ONR, NRI-SRC-SWAN, and
LPS-NSA.

\bibliographystyle{elsartbst-dave}
\bibliography{bibtex-sorted}

\begin{thebibliography}{10}
\expandafter\ifx\csname url\endcsname\relax
  \def\url#1{\texttt{#1}}\fi
\expandafter\ifx\csname urlprefix\endcsname\relax\def\urlprefix{URL }\fi
\expandafter\ifx\csname href\endcsname\relax
  \def\href#1#2{#2} \def\path#1{#1}\fi

\bibitem{abergel-advphys59}
D.~S.~L. Abergel, V.~Apalkov, J.~Berashevich, K.~Ziegler, T.~Chakraborty, Adv.
  Phys. 59 (2010) 261.

\bibitem{dassarma-rmp83}
S.~Das~Sarma, S.~Adam, E.~H. Hwang, E.~Rossi, Rev. Mod. Phys. 83 (2011) 407.

\bibitem{castroneto-rmp81}
A.~H. Castro~Neto, F.~Guinea, N.~M.~R. Peres, K.~S. Novoselov, A.~K. Geim,
  {Rev. Mod. Phys.} {81} ({2009}) {109}.

\bibitem{mccann-prl96}
E.~McCann, V.~I. Fal'ko, {Phys. Rev. Lett.} {96} ({2006}) {086805}.

\bibitem{mccann-prb74}
E.~McCann, Phys. Rev. B 74 (2006) 161403.

\bibitem{kuzmenko-prb80}
A.~B. Kuzmenko, I.~Crassee, D.~van~der Marel, P.~Blake, K.~S. Novoselov, Phys.
  Rev. B 80 (2009) 165406.

\bibitem{fang-apl91}
T.~Fang, A.~Konar, H.~Xing, D.~Jena, Appl. Phys. Lett. 91 (2007) 092109.

\bibitem{abergel-prb80}
D.~S.~L. Abergel, P.~Pietil\"ainen, T.~Chakraborty, Phys. Rev. B 80 (2009)
  081408.

\bibitem{martin-natphys4}
J.~Martin, N.~Akerman, G.~Ulbricht, T.~Lohmann, J.~H. Smet, K.~von Klitzing,
  A.~Yacoby, Nat. Phys. 4 (2008) 144.

\bibitem{hwang-prl99}
E.~H. Hwang, B.~Y.-K. Hu, S.~Das~Sarma, Phys. Rev. Lett. 99 (2007) 226801.

\bibitem{barlas-prl98}
Y.~Barlas, T.~Pereg-Barnea, M.~Polini, R.~Asgari, A.~H. MacDonald, Phys. Rev.
  Lett. 98 (2007) 236601.

\bibitem{kusminskiy-prl100}
S.~V. Kusminskiy, J.~Nilsson, D.~K. Campbell, A.~H. Castro~Neto, Phys. Rev.
  Lett. 100 (2008) 106805.

\bibitem{borghi-prb82}
G.~Borghi, M.~Polini, R.~Asgari, A.~H. MacDonald, Phys. Rev. B 82 (2010)
  155403.

\bibitem{vafek-prb81}
O.~Vafek, K.~Yang, Phys. Rev. B 81 (2010) 041401.

\bibitem{zhang-prb81}
F.~Zhang, H.~Min, M.~Polini, A.~H. MacDonald, Phys. Rev. B 81 (2010) 041402.

\bibitem{martin-prl105}
J.~Martin, B.~E. Feldman, R.~T. Weitz, M.~T. Allen, A.~Yacoby, Phys. Rev. Lett.
  105 (2010) 256806.

\bibitem{oostinga-natmat7}
J.~B. Oostinga, H.~B. Heersche, X.~Liu, A.~F. Morpurgo, L.~M.~K. Vandersypen,
  Nat. Mater. 7 (2008) 151.

\bibitem{asgari-prb77}
R.~Asgari, M.~M. Vazifeh, M.~R. Ramezanali, E.~Davoudi, B.~Tanatar, Phys. Rev.
  B 77 (2008) 125432.

\bibitem{deshpande-prb79}
A.~Deshpande, W.~Bao, F.~Miao, C.~N. Lau, B.~J. LeRoy, Phys. Rev. B 79 (2009)
  205411.

\bibitem{rutter-natphys7}
G.~M. Rutter, S.~Jung, N.~N. Klimov, D.~B. Newell, N.~B. Zhitenev, J.~A.
  Stroscio, Nat. Phys. 7 (2011) 649.

\bibitem{deshpande-apl95}
A.~Deshpande, W.~Bao, Z.~Zhao, C.~N. Lau, B.~J. LeRoy, Appl. Phys. Lett. 95
  (2009) 243502.

\bibitem{rossi-prl101}
E.~Rossi, S.~Das~Sarma, Phys. Rev. Lett. 101 (2008) 166803.

\bibitem{rossi-prl107}
E.~Rossi, S.~Das~Sarma, Phys. Rev. Lett. 107 (2011) 155502.

\bibitem{adam-prb84}
S.~Adam, S.~Jung, N.~N. Klimov, N.~B. Zhitenev, J.~A. Stroscio, M.~D. Stiles,
  Phys. Rev. B 84 (2011) 235421.

\bibitem{gibertini-arXiv1111}
M.~{Gibertini}, A.~{Tomadin}, F.~{Guinea}, M.~I. {Katsnelson}, M.~{Polini},
  ArXiv e-prints arXiv:1111.6280.

\bibitem{guinea-prb77}
F.~Guinea, M.~I. Katsnelson, M.~A.~H. Vozmediano, Phys. Rev. B 77 (2008)
  075422.

\bibitem{abergel-prb83}
D.~S.~L. Abergel, E.~H. Hwang, S.~Das~Sarma, Phys. Rev. B 83 (2011) 085429.

\bibitem{abergel-prb84}
D.~S.~L. Abergel, H.~Min, E.~H. Hwang, S.~Das~Sarma, Phys. Rev. B 84 (2011)
  195423.

\bibitem{burson-unpub}
K.~Burson,  \textit{et al.}, (2011), (unpublished).

\bibitem{henriksen-prb82}
E.~A. Henriksen, J.~P. Eisenstein, Phys. Rev. B 82 (2010) 041412.

\bibitem{young-prb84}
A.~F. Young, L.~S. Levitov, Phys. Rev. B 84 (2011) 085441.

\end{thebibliography}

\end{document}